# Strain coupling optimization in magnetoelectric transducers

D. Tierno [a, b*], F. Ciubotaru [a], R. Duflou [a], M. Heyns [a, b],
I. P. Radu [a], C. Adelmann [a]

*[a] Imec, Kapeldreef 75, 3001 Leuven, Belgium*
*[b] KU Leuven, Departement Materiaalkunde (MTM), B-3001 Leuven, Belgium*

**Abstract**

The mechanical behavior of magnetoelectric transducers consisting of piezoelectric-magnetostrictive bilayers has been modeled. The effect of the aspect ratio of the transducer as well as the influence of non-active surrounding layers has been modeled, including a passivation layer surrounding the active device, a clamping layer above the active device, and an interfacial layer that might be inserted between the magnetostrictive and the piezoelectric layers. Strategies to control and maximize the strain magnitude and orientation based on material selection and device design are proposed.

## 1. Introduction

Current semiconductor-based CMOS devices may reach their physical limits in the next decade. To be able to continue Moore's law, to improve the device performance, and to further lower the power per operation, the replacement of CMOS circuits by novel circuits based on different physical effects may become necessary. In particular, logic circuits based on the interference of spin waves [1, 2] are a promising alternative to CMOS technology and are highly suitable to efficiently implement majority gates [1 - 4], in which the state of the output is determined by the majority of the input states. In such spin wave logic gates, the information is encoded in the phase of the waves and the output is determined by the interference of multiple spin waves propagating in a common waveguide. Spin-wave computing has the potential for low-power computation since no charge motion takes place. Furthermore, it allows the functional scaling of the circuit [1] which would relax the high density of on chip-elements required in the CMOS technology. To build logic circuits based on spin-wave majority gates that are competitive with CMOS-based technology, it is necessary to develop energy efficient transducers between spin-wave and electric domains so to cointegrate the two technologies in a single system. Key requirements of such transducers are high coupling efficiency, low operational power, and high bandwidth [1-3]. Microwave antennae have typically been used to generate spin waves using electric currents [4, 5] but are neither scalable nor energy efficient. By contrast, magnetoelectric transducers represent a scalable and low-power alternative [2, 3] consisting of a piezoelectric-magnetostrictive (PE-MS) bilayer in which the coupling between the electric and the spin domain occurs via strain: when an electric field is applied across the piezoelectric, strain is induced and transferred to the magnetostrictive film that in turn changes its magnetic anisotropy via the inverse magnetostrictive effect. The resulting change of the effective magnetic anisotropy field can exert a torque on the magnetization and, in case of an AC excitation, generate spin waves.

Most of the research performed in this field has focused on the coupling mechanism of magnetoelectric bilayers. By contrast, little interest instead is shown for actual micro- or nanoscale devices. Materials with high piezoelectric and magnetostrictive coefficients are desirable for the PE-MS bilayer but strain transfer optimization within the bilayer in patterned transducers presents several integration challenges. To design efficient micro- and nanomechanical transducers width dimensions of few µm or less, many aspects need to be addressed such as the impact of the geometry of the patterned transducer as well as the mechanical properties of all materials in fully integrated magnetoelectric transducers.

[*] Corresponding author - Davide Tierno: davide.tierno@imec.be;  Kapeldreef 75,  3001 Leuven - Belgium

The stress-strain relation is described but the Generalized Hooke's law:

$$\varepsilon_{xx} = \frac{\sigma_{xx}}{E} - \upsilon\frac{\sigma_{yy}}{E} - \upsilon\frac{\sigma_{zz}}{E} \quad (1)$$

$$\varepsilon_{yy} = -\upsilon\frac{\sigma_{xx}}{E} + \frac{\sigma_{yy}}{E} - \upsilon\frac{\sigma_{zz}}{E} \quad (2)$$

$$\varepsilon_{zz} = -\upsilon\frac{\sigma_{xx}}{E} - \upsilon\frac{\sigma_{yy}}{E} + \frac{\sigma_{zz}}{E} \quad (3)$$

with $E$ the Young's modulus of the material and $\upsilon$ the Poisson ratio.

The magnetoelastic torque $\tau$ on the magnetization due to a strain field is given by [6]:

$$\tau_{mel} = \begin{pmatrix} 2B_1 m_y m_z (\varepsilon_{yy} - \varepsilon_{zz}) + B_2 (m_z m_x \varepsilon_{xy} - m_x m_y \varepsilon_{zx} + (m_z^2 - m_y^2)\varepsilon_{yz}) \\ 2B_1 m_z m_x (\varepsilon_{zz} - \varepsilon_{xx}) + B_2 (m_x m_y \varepsilon_{yz} - m_y m_z \varepsilon_{xy} + (m_x^2 - m_z^2)\varepsilon_{zx}) \\ 2B_1 m_x m_y (\varepsilon_{xx} - \varepsilon_{yy}) + B_2 (m_y m_z \varepsilon_{zx} - m_z m_x \varepsilon_{yz} + (m_y^2 - m_x^2)\varepsilon_{xy}) \end{pmatrix} \quad (4)$$

where $m_{x,y,z} = M_{x,y,z}/M_S$ are the normalized components of the magnetization vector $M$ with respect the saturation magnetization $M_S$, $\varepsilon_{i,j}$ are the components of the strain tensor within the magnetostrictive layer, $B_1$ and $B_2$ are the magnetoelastic coupling constants, and $\tau$ is the exerted torque, $H$ is the effective magnetic field associated with the magnetoelasticity and $\mu_0$ the permeability of vacuum.

From Eq. 4 it is obvious that $\varepsilon_{i,j}$ plays an important role on the magnitude of the torque exerted on the magnetization due to the magnetoelectric effect, and implicitly, on the excitation efficiency of spin waves. Therefore, to develop a suitable model of the transducer to define strategies and guidelines to control the magnitude and the distribution of $\varepsilon_{i,j}$ is of crucial interest. In this work, we proposed a design of a such magnetoelectric transducer and we analyzed the behavior of the strain tensor components $\varepsilon_{i,j}$ as a function of the geometrical parameters. Further, we studied the influence of additional features of the design as the presence of a passivation, interlayer or and the impact of their mechanical properties on the response of the device.

## 2. Model Description

To quantitatively assess the magnetoelastic and magnetoelectric coupling in magnetoelectric transducers, the mechanical response of a rectangular cuboid transducer was studied using COMSOL Multiphysics simulations.

Of particular interest was the geometry of the transducer and the impact of the non-active surrounding layers on the effective magnetoelastic and magnetoelectric coupling. A schematic of the transducer and its cross section are shown Figs. 1A and 1B, respectively. Such a geometry is perhaps the easiest to fabricate and has been proposed in several publications [7-11]. The "active" part of the structure consisted of a pillar including the magnetoelectric bilayer and two metal electrodes. Underneath, a 400 nm thick $SiO_2$ layer provided electrical insulation from the 20 μm Si substrate. The model also included an extended free area around the pillar (see Fig. 1A) so that the edges of the pillar can displace freely. This free region was set to be five times the width (W) of the pillar and scaled accordingly. The bottom face of the Si box was set as fixed in the boundary conditions of the model, which thus includes potential deformations of the substrate.

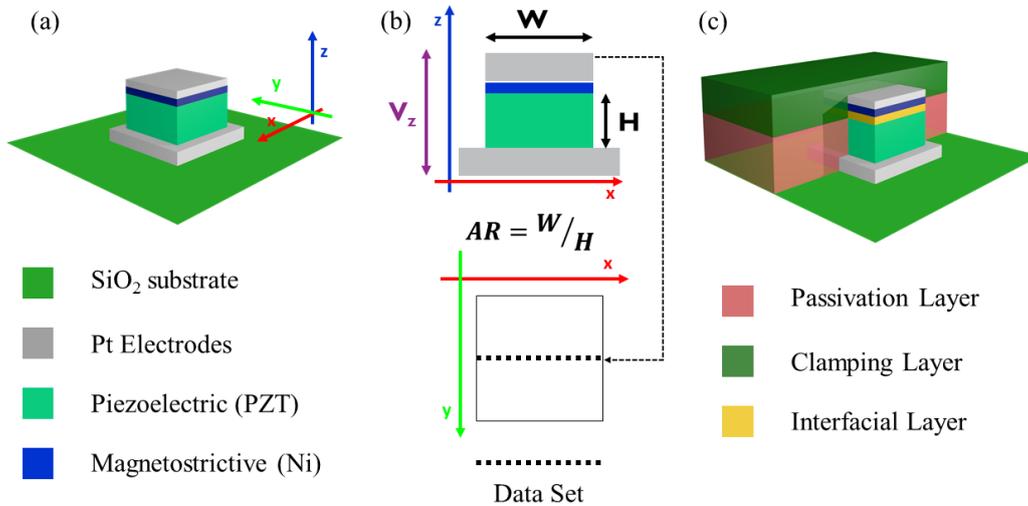

Figure 1 - (a) Schematic and (b) Cross Section of the rectangular cuboid magnetoelectric transducer; (c) Schematic of the magnetoelectric transducer including the non-active layers.

Successively, three non-active layers were added, as shown in Fig. 1C: a passivation layer surrounding the active transducers, a clamping layer above the transducer, and an interfacial layer that might be inserted between the magnetostrictive and the piezoelectric layers. The three non-active layers were added to the model one at the time to evaluate their impact individually. The range of Young's moduli used in the simulations for the three layer was defined after characterizing a set of candidate materials by nanoindentation [12, 13].

Table 1 shows a more detailed summary of the geometric dimensions and the materials used in the model. Below, we quantitatively assess the variation of the magnetoelastic coupling across the different simulated scenarios with the purpose to define strategies and guidelines to optimize and engineer the strain coupling.

|  | **Symbol/Formula** | **Value (or Range)** | **Material** |
|---|---|---|---|
| Width Transducer | $W$ | 20 nm – 20 µm | - |
| Thickness Magnetostrictive | $tms$ | 10 nm | Nickel |
| Thickness Piezoelectric | $H$ | 100 nm – 200 nm | PZT |
| Thickness Substrate | $tsub$ | 20 um | Si |
| Thickness SiO$_2$ | $tox$ | 400 nm | SiO$_2$ |
| Thickness Electrodes | $tel$ | 70 nm | Pt |
| Extension Free Area | $5 \times W$ | 100 nm – 100 µm | - |
| Thickness Passivation Layer | $tpass\ (= tpe)$ | 100 nm – 200 nm | - |
| Thickness Clamping Layer | $tclamp$ | 200 nm | - |
| Thickness Interfacial Layer | $tint$ | 10 nm | - |

Table 1 – Geometric dimensions and materials of the active and non-active elements of the magnetoelectric transducer.

### 3. Results and discussion

#### 3.1 Aspect Ratio

We first discuss the behavior of the active transducer pillar without additional layers to evaluate the impact of the aspect ratio on the displacement of the piezoelectric layer and eventually on the coupling efficiency. Keeping the thickness H of the piezoelectric layer constant at 200 nm, the aspect ratio AR = W/H was varied between 0.1 and 100 by progressively increasing the pillar width W from 20 nm to 20 µm. The largest aspect ratios can be considered to mimic blanket layers. In all simulations below, an electric field of 50 kV/cm was applied across the magnetoelectric bilayer. The in-plane and out-of-plane displacement components at the PE-MS interface are plotted vs. the normalized length for different aspect ratios Figs. 2A and 2B, respectively.

The maximum in-plane displacement was localized at the edges for blanket layers and decreased significantly with the aspect ratio, particularly for AR < 5. The opposite was observed for the out-of-plane displacement that was three times larger in small pillars than in blanket layers. Thus, in pillar-like structures the out-of-plane displacement was predominant while in blanket layers the lateral displacement dominated. The difference can be related to the contribution of substrate clamping to the displacement of the piezoelectric element: blanket layers are more affected than pillar-like structures due to the significantly larger footprint. Similar results have been previously described in literature [14]. The strain induced in the magnetostrictive layer followed the same trend as shown in Fig. 2C: $\varepsilon_{zz}$ increased significantly while $\varepsilon_{xx}\ (= \varepsilon_{yy})$ decreased when the AR of the transducer was lower than 5.

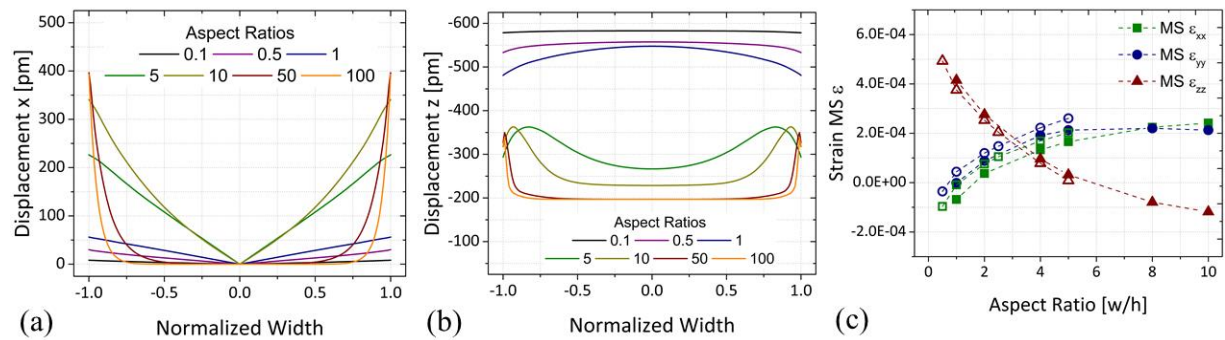

Figure 2 – (a) x- and (b) z-component of the displacement at the PE-MS interface for different aspect ratios (AR = W/H). (c) Impact of the Young's Modulus of the Passivation Layer on $\varepsilon_{xx}$, $\varepsilon_{yy}$, and $\varepsilon_{zz}$ evaluated in the magnetostrictive layer.

### 3.2 Passivation Layer

The purpose of the passivation layer is the isolation of different magnetoelectric cells from each other; moreover, it provides mechanical support for waveguides in full spin wave devices. The key material parameter of the passivation layer is its Young's modulus and thus several materials ranging from to very stiff materials (such as SiN) to less stiff materials (such as spin-on-glass, SOG, or spin-on-carbon, SoC) were included in the simulation. The passivation layer was thus added to the model and its Yong's modulus was varied from 1 GPa (SoC/SOG) to 200 GPa (SiN). The evaluated values of $\varepsilon_{xx}$, $\varepsilon_{yy}$, and $\varepsilon_{zz}$ are shown in Fig. 3A.

Increasing the stiffness of the passivation layer led to a strong reduction of the in-plane strain components $\varepsilon_{xx}$ and $\varepsilon_{yy}$ although they remained positive. On the other hand, $\varepsilon_{zz}$ decreased in magnitude between 1 and 50 GPa. However, for higher Young's moduli, it switched from compressive to tensile strain and increased again. At 1 GPa and at 200 GPa, the out-of-plane strain components had the same magnitude but with opposite direction. Such behavior can be explained by the limitations imposed by a stiff passivation layer on the piezoelectric element, reducing the freedom for lateral displacement with respect to a free-standing pillar. To avoid the reduction of the $\varepsilon_{xx}$ and $\varepsilon_{yy}$, it is therefore advisable to use a passivation layer with a very low Young's modulus. On the other hand, it is interesting to notice that not only the magnitude but also the direction of the strain components can be controlled by varying the stiffness of the Passivation Layer.

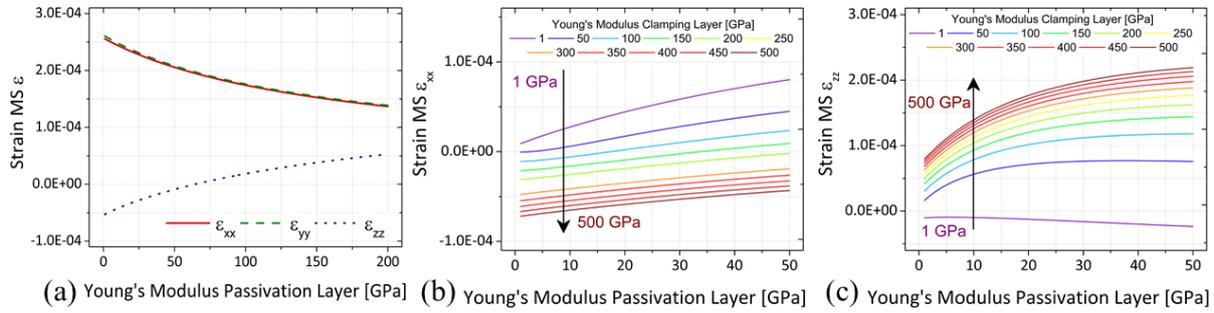

Figure 3 – (a) Impact of the Young's Modulus of the Passivation Layer on the main strain components in the MS layer. (b) $\varepsilon_{xx}$ and (c) $\varepsilon_{zz}$ for different combinations of Passivation (x-axis) and Clamping Layers (color code in the legend).

### 3.3 Clamping Layer

An integrated electronic circuit typically consists of different device levels on top of each other and separated by a passivation layer. The same scenario was considered for integrated magnetoelectric transducers. in our model a 200nm-thick layer was added to the model defined as the clamping layer in Fig. 1C. The first point to address is whether the same material should be used for both the passivation and the clamping layer. In our simulations, the Young's modulus of the clamping layer was varied between 1 and 500 GPa, while independently varying the Young's modulus of the passivation layer between 1 to 50 GPa. The clamping layer limits the displacement of the magnetoelectric transducer that cannot expand neither vertically nor laterally. By reducing the possibility of expansion, the force exerted by the displacement of the piezoelectric element leads to a much larger pressure exerted on the magnetostrictive layer in the pillar. This limitation had a positive impact on the magnetoelastic coupling, as shown by the strong increase of both the in-plane and out-of-plane strain components, irrespective of the Young's modulus of the passivation layer, in Figs. 3B and 3C, respectively. The in-plane strain became compressive if the Young's modulus of the clamping layer was larger than 100 GPa and its magnitude increased by increasing the stiffness of the layer. Consequently, $\varepsilon_{zz}$ increased significantly, up to one order of magnitude, for extremely compliant passivation layers. For stiffer clamping layers, the increase was less significant. Thus, combining a compliant passivation layer and a stiff clamping layer can lead to a tenfold increase of $\varepsilon_{zz}$ without reducing $\varepsilon_{xx}$ and $\varepsilon_{yy}$. By contrast, the in-plane components benefited from a stiff passivation layer as their magnitude increased while the direction of the strain changed from tensile to compressive.

### 3.4 Interfacial Layer

In a third step, a thin metallic layer was introduced between the piezoelectric and the magnetostrictive layer. Such a layer may play an important role in actual magnetoelectric transducers as it might serve e.g. as a seed layer for the magnetostrictive layer. In this set of simulations, the Young's moduli of the passivation and the clamping layers were kept constant to 5 and 200 GPa, respectively. The variation of $\varepsilon_{xx}$ and $\varepsilon_{zz}$ is shown in Fig. 4A. Results of simulations without interfacial and layer are shown for comparison also.

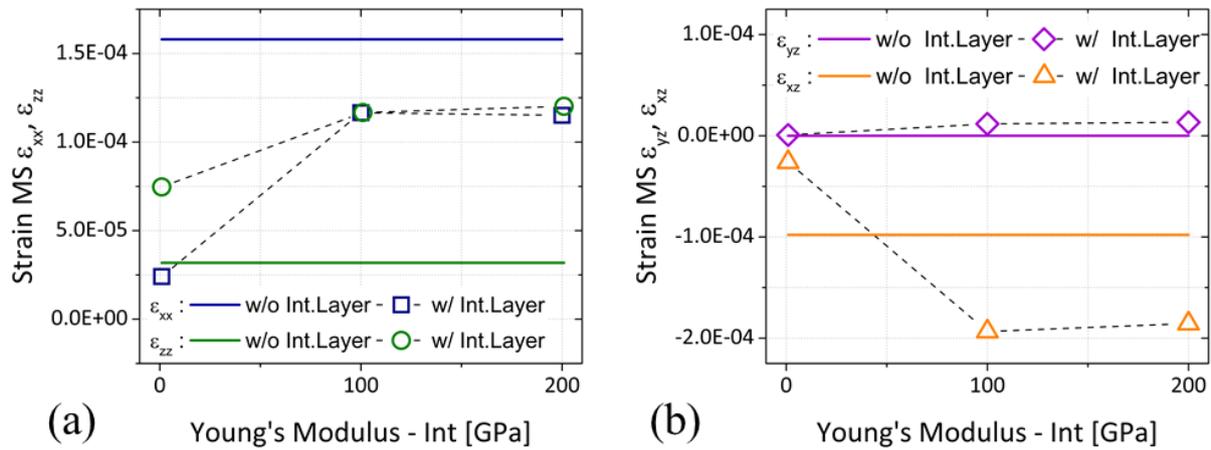

Figure 4 – Impact of the Interfacial Layer (Int. Layer) on (a) $\varepsilon_{zz}$ and $\varepsilon_{xx}$ and (b) $\varepsilon_{xz}$ and $\varepsilon_{yz}$

For a sufficiently stiff interfacial layer with a Young's modulus of 200 GPa (e.g. Ta), a decrease of $\varepsilon_{xx}$ by about one third was observed while leading to an about threefold increase in $\varepsilon_{zz}$. The most important consequence on the introduction of the interfacial layer however was the large enhancement of the shear-strain components $\varepsilon_{(ij, i \neq j)}$, as shown in Fig. 4B. It has been shown by micromagnetic simulations that both biaxial strain as well as shear strain allow for a significantly enhanced magnetoelastic coupling in scaled magnetic waveguides [15]. Hence, guidelines how to optimize biaxial and shear strains may lead to a strongly increased spin wave generation efficiency by magnetoelectric transducers.

## 4. Conclusions

The design and integration of magnetoelectric transducers pose many challenges besides the choice and the optimization of the piezoelectric-magnetostrictive bilayer. The simulations showed that the geometry of the transducer as well as the mechanical properties of surrounding non-active layers play a significant role in the coupling efficiency, regardless of the materials used for the bilayer. In particular, tall pillars showed very different behavior than flat blanket layers, with the out-of-plane strain being the dominating component. Moreover, in-plane and out-of-plane components could be decoupled and controlled separately, not only their magnitude but also the direction of the components. This is a great opportunity for device designers since it gives the opportunity to engineer the response of the transducer according to the required operational conditions since the torque on the magnetization in the magnetostrictive film depends both on the orientation of the magnetization vector as well as on the form of the strain tensor.

**Funding**
This work was supported by imec's Industrial Affiliation Program on Beyond CMOS devices.